\documentclass[%
 reprint,
 superscriptaddress,
 amsmath,amssymb,
 aps,
]{revtex4-2}

\usepackage{graphicx}
\usepackage{dcolumn}
\usepackage{bm}

\usepackage{hyperref}
\hypersetup{
	colorlinks=true,
	citecolor=blue,
	linkcolor=red,
	urlcolor=blue}

\begin{document}


\title{Predicting deformation mechanisms in architected metamaterials using GNN}

\author{Padmeya Prashant Indurkar}
\affiliation{Department of Engineering, University of Cambridge, Cambridge, CB2 1PZ, U.K.}%
\author{Sri Karlapati}%
\thanks{Work done outside of Amazon Research through an informal collaboration.}%
\affiliation{Department of Engineering, University of Cambridge, Cambridge, CB2 1PZ, U.K.}%
\affiliation{Amazon Research, Cambridge, U.K.}%
\author{Angkur Jyoti Dipanka Shaikeea}%
\affiliation{Department of Engineering, University of Cambridge, Cambridge, CB2 1PZ, U.K.}%
\author{Vikram S. Deshpande}%
\affiliation{Department of Engineering, University of Cambridge, Cambridge, CB2 1PZ, U.K.}%

\begin{abstract}

The present paradigm in design and modelling of lattice architected mechanical metamaterials is mostly limited to traditional numerical methods like finite element analysis.
Recently, the use of machine learning and artificial intelligence techniques have become popular and here we extend these ideas to architected metamaterials. We show that truss based lattices have a natural resemblance to computational graphs which serve as an input for the rapidly emerging field of graph neural networks (GNNs). A dataset comprising thousands of such extremely complex lattices is trained using a GNN to predict the underlying dominant deformation mechanism viz. stretching and bending. The trained GNN achieves $>90\%$ accuracy on a previously unseen complex lattice dataset. Such graph-based learning of metamaterials has the capability to predict a range of properties, from elastic moduli to fracture toughness and promises AI driven discovery of emergent metamaterials possessing superlative properties.

\end{abstract}

\maketitle


\section{\label{sec:level1}Introduction}

The emergence of additive manufacturing (AM) techniques has facilitated the fabrication of complex 3D shapes with topological feature sizes spanning length scales from nanometres and upwards \cite{zheng2016multiscale}. These manufacturing technologies have enabled the creation of new materials (metamaterials) with previously unattainable properties resulting from their topology, such as recoverable ceramics, super elastic metallic glasses, and strong and energy absorbing materials with promising applications ranging from space structures to biological implants\cite{surjadi2019mechanical}. Progress in lattice materials has been driven by both, development of new unit cell topologies, or by manufacturing advances that leverage unique materials and nano-scale effects to maximize performance. However, most modelling approaches for lattice materials to-date rely on classical modelling methods like FE \cite{fleck2010}.

Data-driven approaches based on machine learning (ML) have significantly accelerated the discovery of new materials. ML algorithms such as logistic regression \cite{zhu1997_logis}, support vector machines \cite{cortes1995_svm}, decision trees \cite{ho1998_dtree}, boosting \cite{friedman2001_boost} and convolution neural networks \cite{lecun1998_cnn} have been utilized to predict material properties, where the material input features tabulated as data or images are employed for training \cite{gu2018_mathor, bessa2019_advmat, wilt2020_advengmat, kumar2020_compmat, mao2020_sciadv}. These approaches largely rely on the application of existing ML models, e.g., deep learning for image classifications, which efficiently capture the hidden patterns of a euclidean dataset. Most present ML approaches however are not tailored to the three-dimensional structure of metamaterials leading to their poor input featurization.

In this regard, there are several emerging applications where data takes a non-euclidean form, i.e.~computational graphs. These include, the interactions between users and products for e-commerce product recommendations; identification of the properties of molecules modelled as graphs for drug discovery; categorization of journal papers into subject areas by linked citations, among others. In applications such as these, geometric deep learning is conducted using graph neural networks (GNN) where the input is represented as a graph through its vertices connected via a set of edges possessing different vertex and edge attributes. GNNs can therefore completely represent an arbitrary 3D metamaterial, including its length scale, hierarchy, and material attributes and have a strong physical analogy with the lattice topology.

\begin{figure}[!h]
	\centering
	\includegraphics[width=0.3\textwidth]{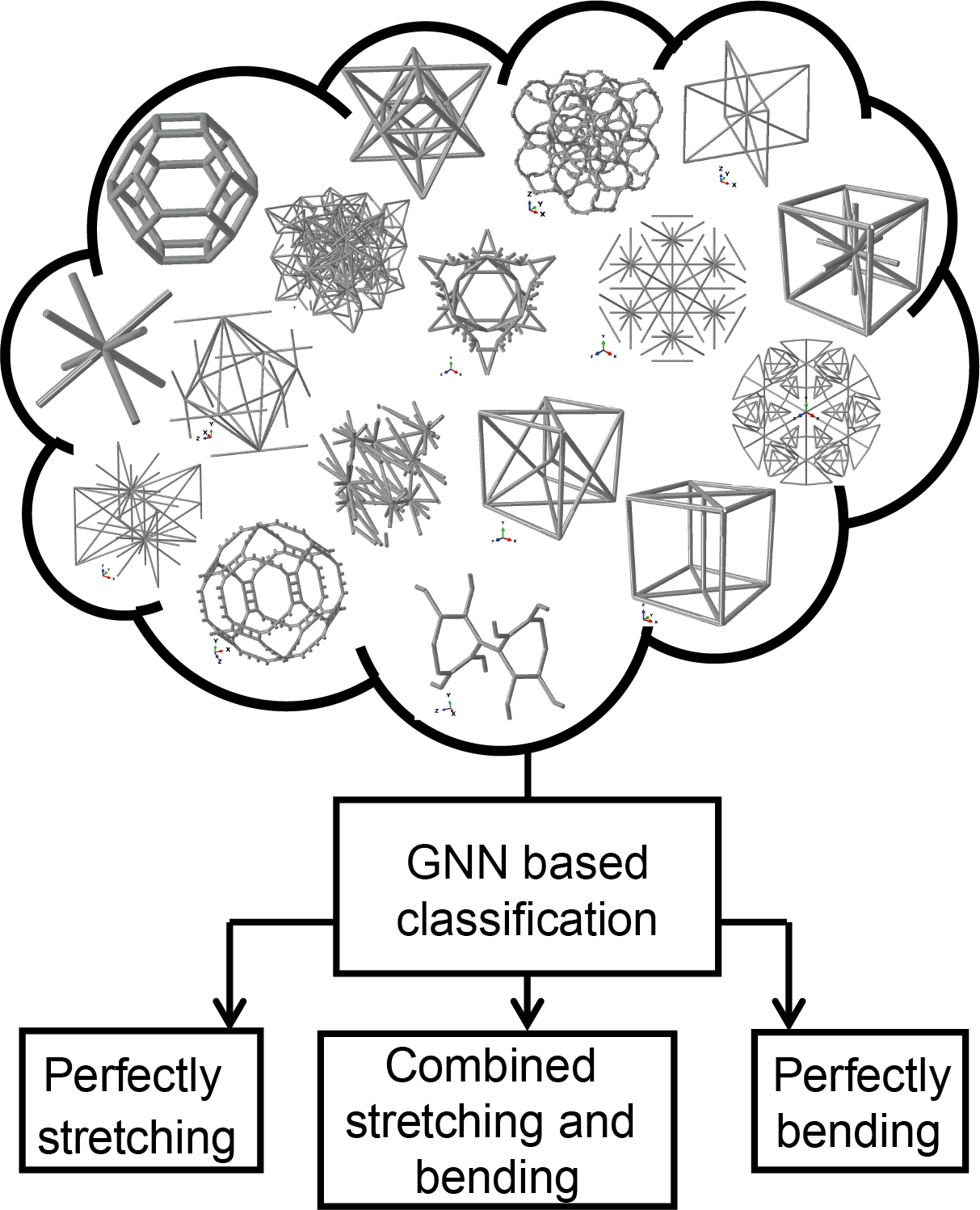}
	\caption{
	\textbf{Scope of the study:} Employing a GNN based classification algorithm to classify a diverse dataset of 3D lattice architected materials into their dominating modes of deformation, viz. stretching, bending and the ones exhibiting combined modes.
	\label{f:fig1}}
\end{figure}

Such GNNs have been efficiently applied in the past to predict chemical properties of molecules, extensively reviewed in \cite{wieder2020}, and in electro-magnetics of polycrystalline alloys. Leveraging these foundational advances in geometric deep learning, in this work we employ a GNN based classifier algorithm for predicting the stretching vs bending behavior of lattice architected materials (cf.~Fig.~\ref{f:fig1}). The dataset employed in this study is described next.

\section{\label{sec:level2}A complex lattice dataset}

The dataset we employ for training the GNN comprises of 17,201 unique 3D lattice topologies adopted from an open-source catalogue \cite{lumpe_pnas_2021}. The lattices are periodic in space and span a wide range of connectivity and crystallographic symmetries (namely cubic, hexagonal, tetragonal, trigonal, orthorhombic, and monoclinic). Along with the nodal coordinates, and an edge connectivity list, the catalogue also provides the elastic properties of lattices, defined next.

\begin{figure}[!htb]
	\centering
	\includegraphics[width=0.45\textwidth]{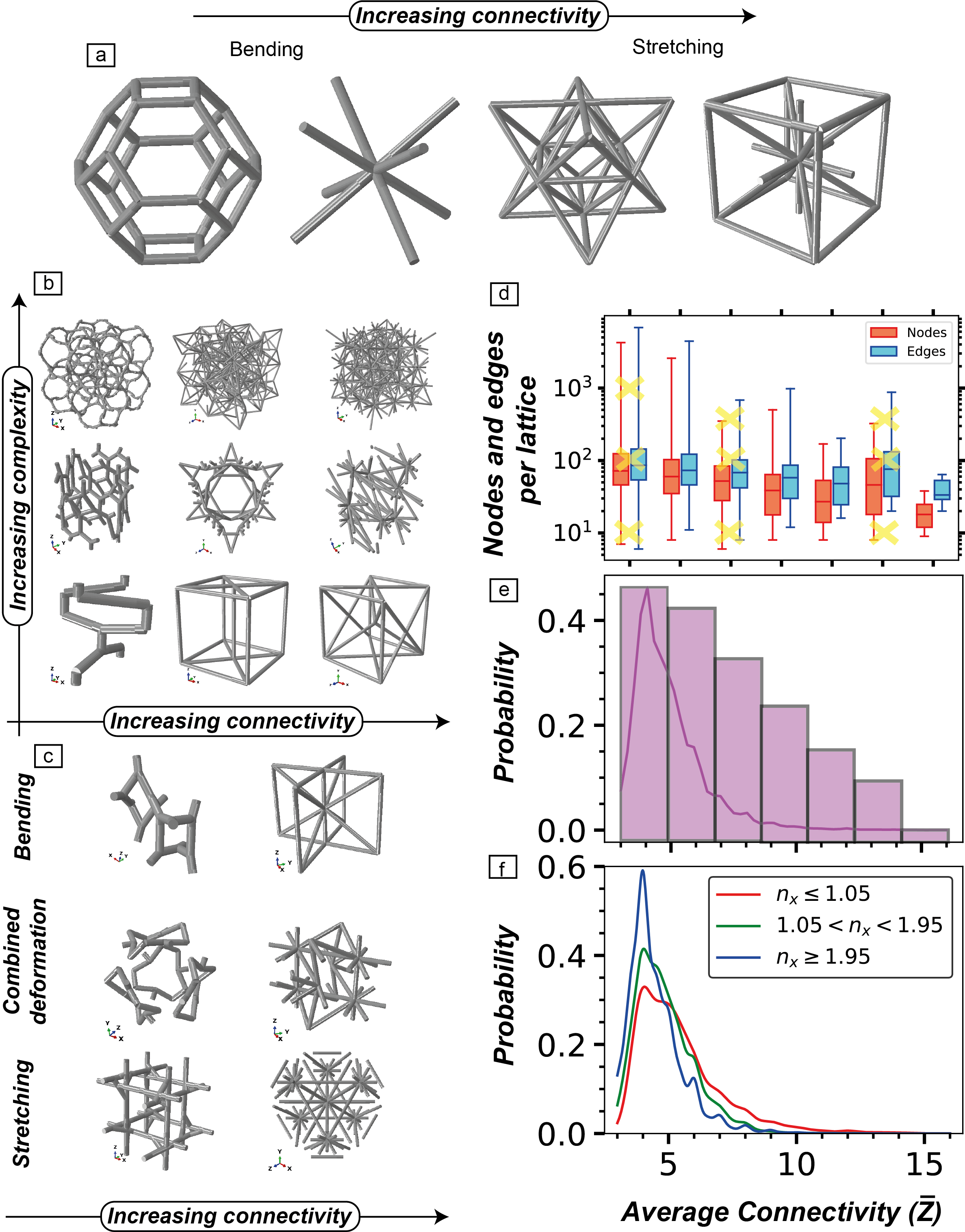}
	\caption{
		\textbf{A diverse lattice dataset:} (a) some of the well-known stretching and bending lattices, viz, kelvin foam, BCC, Octet and reinforced-BCC from the dataset. (b) The different lattices exhibiting varying levels of average connectivity ($\bar{Z}$) and complexity (governed by the total number of nodes and edges per lattice). (c) The lattices exhibiting varying levels of $\bar{Z}$ with different dominating mechanisms. (f) The statistics of nodes and edges per lattice with $\bar{Z}$ for the bins shown in panel (e). Panel (e) shows the probability distributions of all the lattices as a function of $\bar{Z}$ and panel (f) show the same for three different classes of lattices.
		\label{f:fig2}}
\end{figure}

Let the lattices have a unit cell length $\ell$ and be made from a parent material of Young’s modulus $E_{S}$(=1MPa) and Poisson’s ratio $\nu_{S}$(=0.3). Let $\bar{\rho}\ (=0.01)$ denote the relative density (volume of solid material to volume of the effective smeared-out continuum) of a given lattice topology controlled using the strut radius, $r$. The linear-elastic effective properties of all lattice topologies along principal loading directions are provided in the catalogue using a numerical homogenization approach \cite{cheng2013}. In 3D, six independent load cases are required to determine the homogenized effective stiffness matrix, which can be expressed using the Young’s moduli, the shear moduli and the along the principal directions, namely $E_{x}$, $E_{y}$, $E_{z}$, and $G_{yz}$, $G_{xz}$, $G_{xy}$, respectively, and Poisson’s ratios. The scaling relationship between the relative density and the effective moduli is of the form
\begin{equation}
    E_i = E_S C_{i} {\bar{\rho}}^{n_{i}}
\end{equation}
where $C_i$ is a topology dependent constant, and $n_i$ is the scaling exponent that governs the deformation mechanism along $i^{\text{th}}$ direction, $i \in (x,y,z)$.
The lattice has a stretching dominated behaviour for $n_{i} \sim 1$, bending dominated for $n_{i} \sim 2$ and combined deformation modes for $1<n_{i}<2$. The effective elastic moduli $E_i$ span several orders of magnitude between ${10}^{-8}$ and ${10}^{-3}$ in the range of lattices considered in the catalogue. This is due to the lattices having widely varying topology, connectivity and number of nodes and edges as evidenced in Fig.~\ref{f:fig2}.

While it is known that the dominant deformation mechanism (stretching vs bending) depends primarily on the average lattice connectivity denoted here by $\bar{Z}$ \cite{deshpande2001}, it can be influenced by the geometrical and topological features of the lattice. Thus, predicting the dominant mode of deformation along a given loading for a lattice structure is a complicated problem here, we solve it using a data-driven deep learning approach employing GNNs. GNNs allow a definition of the lattice structure using its nodes, edges, their features, and its connectivity. The representation of the lattice is described next.

\begin{figure*}[!htb]
	\centering
	\includegraphics[width=1\textwidth]{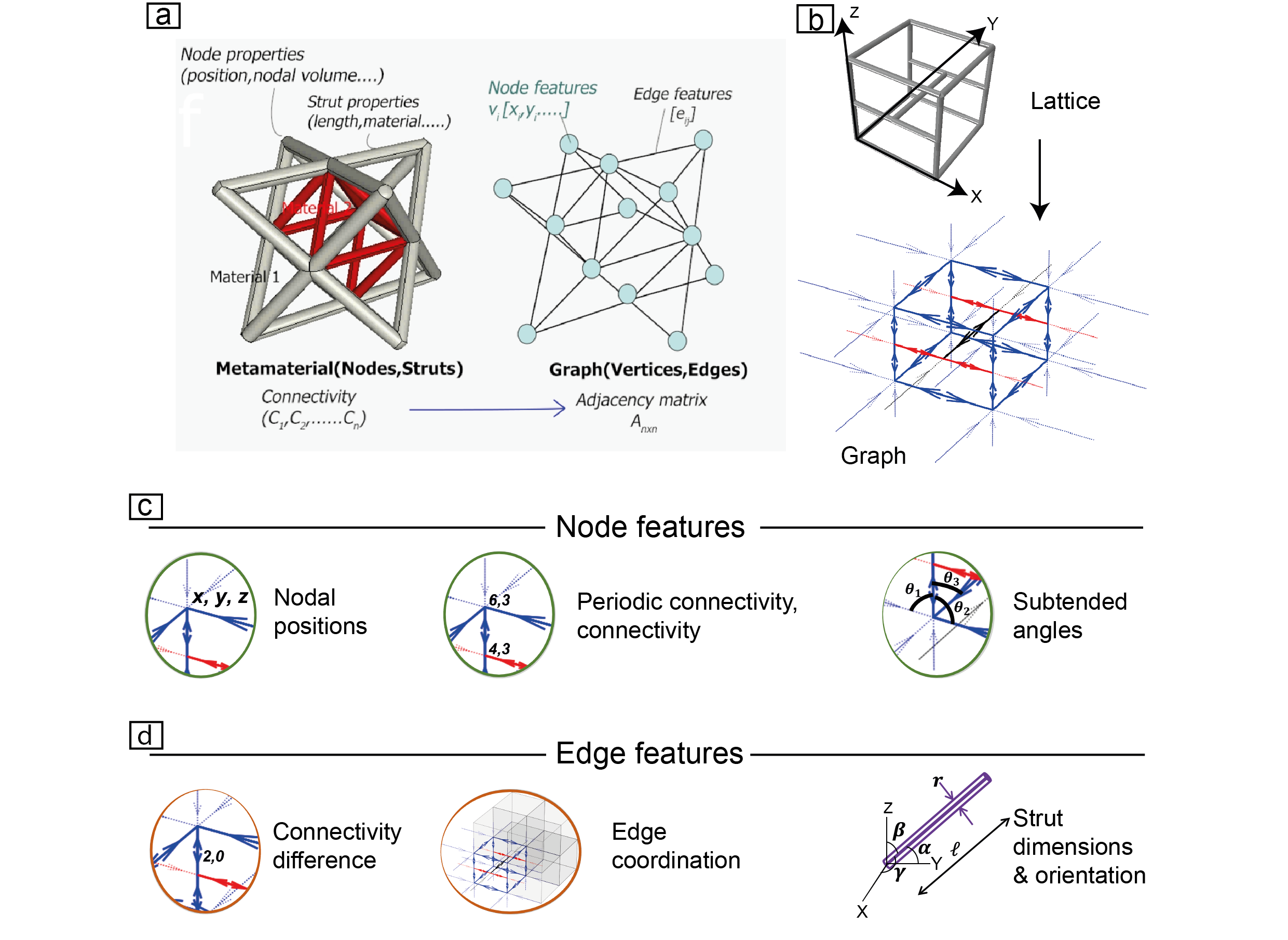}
	\caption{
		\textbf{Lattice to graph representation:} (a) Representation of a lattice as a computational graph required as an input for the GNN shown with its nodes, edges, and the corresponding features as well as connectivity adjacency matrix. (b) The computational graph with message passing directions for a sample lattice belonging to the dataset. (c) and (d) show the different node and edge features adopted, respectively.
		\label{f:fig3}}
\end{figure*}

\section{Lattice to graph representation}

A lattice is defined using a graph $\mathcal{G}=(\mathcal{V},\mathcal{E})$ where $\mathcal{V}$ and $\mathcal{E}$ respectively refer to a set of $n$ vertices (or nodes) and $m$ edges (or struts) constituting the lattice. The $i^{\text{th}}$ vertex in a set of $\mathcal{V}$ vertices is denoted by $v_i$ and has a $N$ dimensional feature vector given by $v_i^F \in \mathcal{R}^{n\ \times\ N}$. Similarly, an edge belonging to the set of $\mathcal{E}$ edges pointing from the $i^{\text{th}}$ vertex to the $j^{\text{th}}$ vertex is denoted by $e_{ij}$, and has a $M$ dimensional feature vector $e_{ij}^F \in \mathcal{R}^{m\ \times\ M}$. The connectivity is provided using the $n \times n$ adjacency matrix $A$ such that $A_{ij}=1$ if $e_{ij} \in \mathcal{E}$ and $A_{ij}=0$ if $e_{ij}  \notin \mathcal{E}$. This is illustrated in Fig.~\ref{f:fig3}a.

Given a base lattice, we first employ geometric transformations to make the lattice invariant to translation and scaling. For all the edges belonging to the unit cell, we add reversed edge connections. This makes the edge description invariant to ordering. Further, on the transformed lattice unit cells, the following modification is performed to have an influence of the translational periodicity of the lattices. We connect the nodes on the outer peripheral faces of the lattice unit cell to the nodes in the neighbouring unit cells arising from periodicity (cf.~Fig.~\ref{f:fig3}b). 

The next step in setting up the GNN is to define the node feature vector $v_i^F$ of each node within the lattice. The possible node features are the normalized nodal coordinates (i.e., $\bar{v}_x=v_x/\ell$, $\bar{v}_y=v_y/\ell$ and $\bar{v}_z=v_z/\ell$). The connectivity at each node is the key node feature governing stretching-vs-bending behavior. The connectivity can be expressed assuming periodicity and in a non-periodic setting. Additionally, let the angles subtended by the different edges meeting at a node be given by a vector $\Theta=[\theta_1,\ \theta_2,\ \theta_3,\ldots]$. If $e_v$ be the number of edges meeting at a given vertex $v$, the size of the vector $\Theta$, i.e., the number of angles subtended at the node by all combinations of edges is $\frac{e_v!}{\left(e_v-2\right)!2!}$. We incorporate a maximum, minimum, average and standard deviation of these angles as the node features. These node features are shown in Fig.~\ref{f:fig3}c.

With the nodal features in place, we describe the edge through its features (cf.~Fig.~\ref{f:fig3}d). Firstly, we adopt the dimensional features of the edge, i.e., length ($l$), radius ($r$) and the directional cosines with the $x$, $y$ and $z$ axis, i.e., $\alpha$, $\beta$ and $\gamma$ as its features. Next, the edge coordination number, $Z_e$ which defines the number of unit cells shared by an edge as another feature. Additionally, we add a difference of periodic connectivity across the nodes belonging to an edge as an edge feature.

\begin{figure*}[htb]
	\centering
	\includegraphics[width=0.9\textwidth]{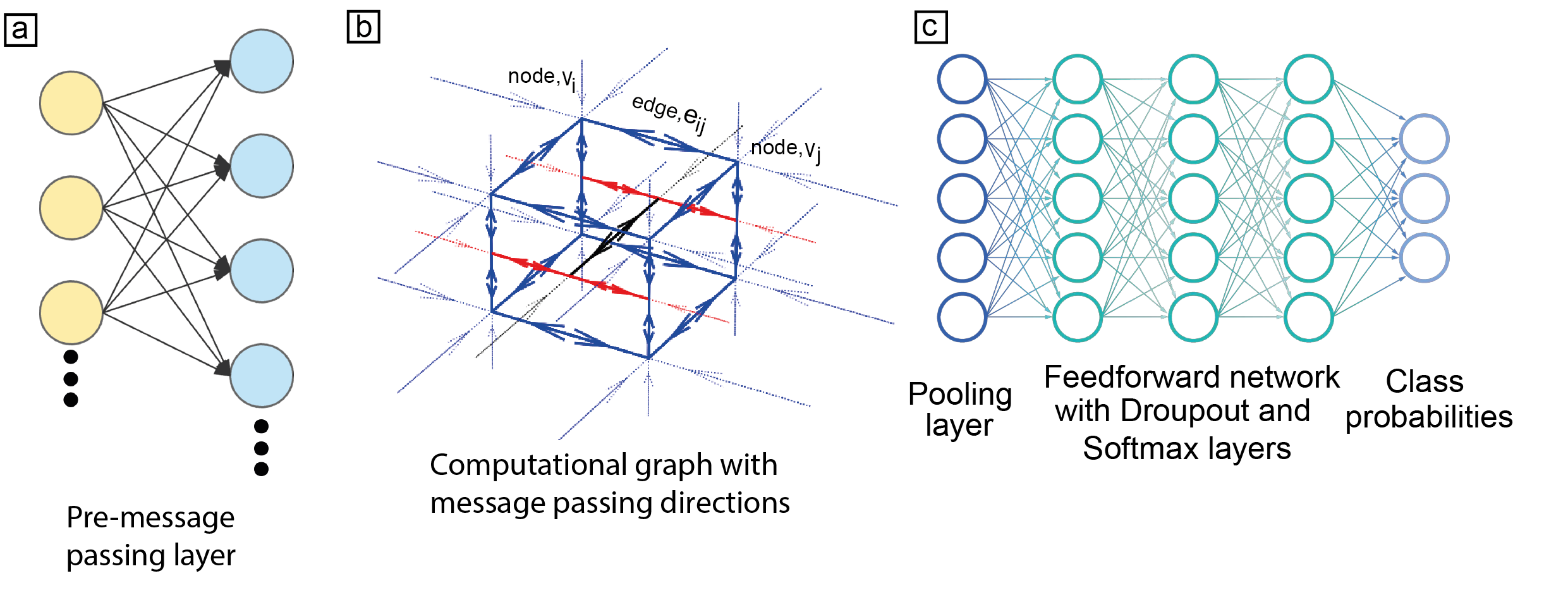}
	\caption{
	\textbf{GNN based classifier architecture:}
	(a) shows the pre-message passing layer which takes the node feature vector and expands it to a higher dimension $d$. (b) shows the message passing in a sample lattice based on the expanded node feature vector. (c) shows the pooling layer followed by feed forward network involving Dropout and Softmax layers resulting in the three class probabilites.
	\label{f:fig4}}
\end{figure*}

\section{THE GNN architecture}

Here, we adopt the message passing GNN architecture shown schematically in Fig.~\ref{f:fig4} as developed in \cite{gilmer2017neural}. The forward pass of the GNN has three phases, a message passing phase, an update phase and a global readout phase. The message passing phase runs for $t=1$ to $T$ time steps and is defined in terms of message function followed by an update phase involving a vertex update function.

The initial nodal states at node $i$, $h_i^0$ are the initial node features of each node, $v_i^F$. A pre-GNN linear neural network layer is employed to map the features to a larger dimension, $d$. Using this enhanced layer, messages are passed and aggregated at node $i$ between connected edges from its neighbourhood $\mathcal{N} \left(i\right) = \left\{j \in \mathcal{V}\ |\ \left(i,j\right)\ \in\ \mathcal{E}\right\}$ as
\begin{equation}
m_i^{t+1} = \sum_{j\ \in\ \mathcal{N}(i)}{M^t \left(h_i^t, h_j^t, e_{ij}^t\right)}
\end{equation}

Using this message vector, the node state is updated using
\begin{equation}
h_i^{t+1} = U^t \left(h_i^t,\ m_i^{t+1}\right)
\end{equation}
After $T$ message passes, a global pooling function $g$ of all the node states belonging to a graph $\mathcal{G}$ results in a layer given by

\begin{equation}
\widetilde{y} = g (\left\{h_i^T\middle| i\ \in\ \mathcal{G}\right\})
\end{equation}

Using the pooled layer, with a feed-forward neural network involving Dropout and SoftMax layers, we reduce the $\widetilde{y}$ vector to a 3-dimensional array representing the probability of the lattice belonging to the three classes, namely perfectly stretching ($p^s$), perfectly bending ($p^b$) and having combined stretching and bending modes ($p^c$). The chosen class is given by the one having the maximum probability. 

Supervised learning to map the actual class with the class predicted by the GNN model involves learning differentiable functions $M^t$ and $U^t$ by minimizing a Cross-Entropy loss. These are chosen as convolution and GRU layers respectively motivated by similar applications in molecular property prediction \cite{gilmer2017neural}. The results from the trained classifier are shown next.

\section{Results and discussion}

We employ all the 17,201 lattices \cite{lumpe_pnas_2021} and classify their linear elastic response along $x$ given by a scaling exponent $n_x$ into (i) perfectly stretching (i.e.~$n_x<1.05$), (ii) perfectly bending (i.e.~$n_x>1.95$) and (iii) the ones exhibiting combined modes. The full dataset was split randomly into 90:10 as training and validation datasets. The training dataset comprises of 15,480 lattices and was further divided into mini-batches of ~100 lattices with random shuffling and the training was performed over ~2,000 epochs using an Adam optimizer. 

\begin{figure*}[!htb]
	\centering
	\includegraphics[width=1\textwidth]{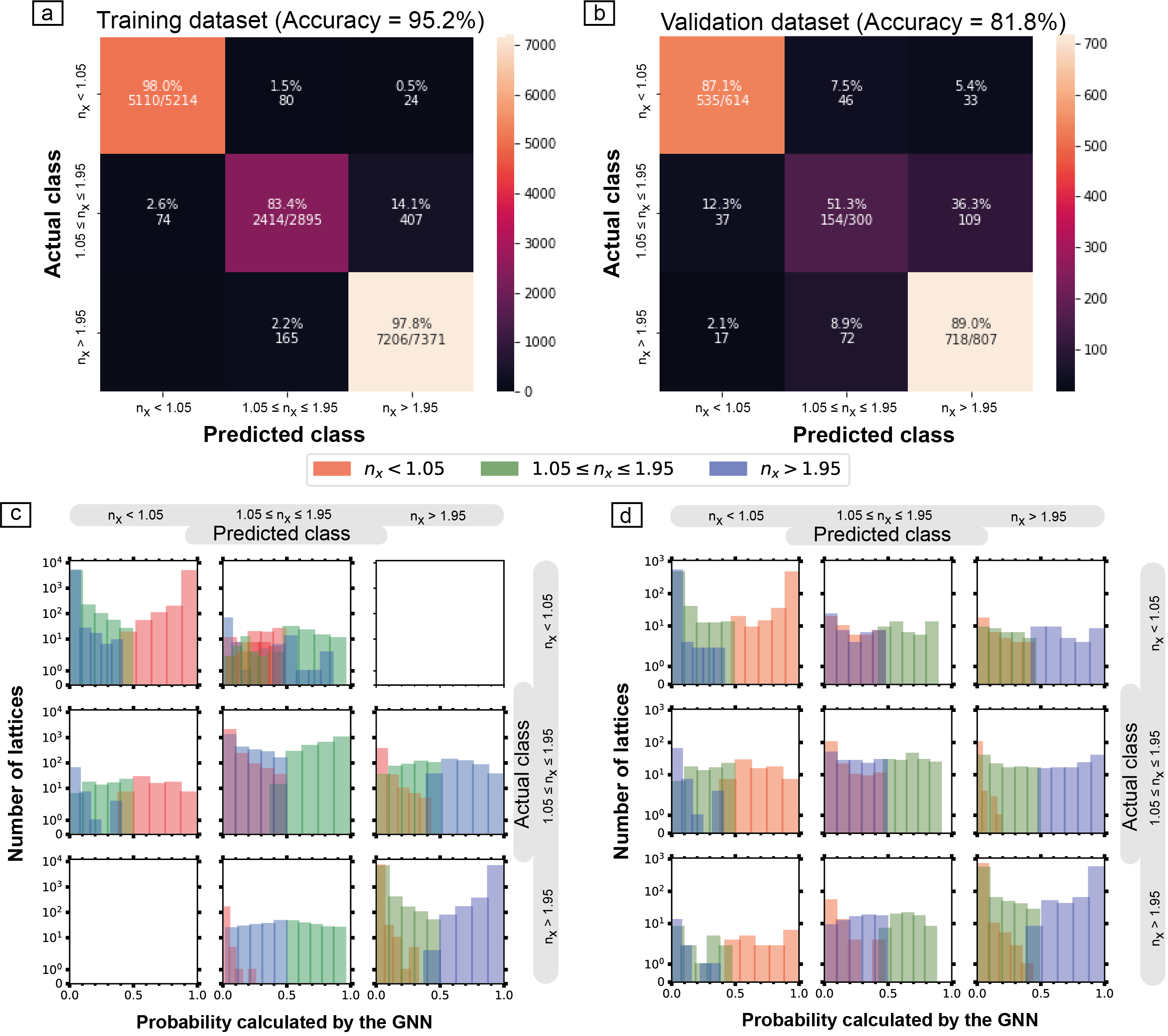}
	\caption{
		\textbf{Results from the trained GNN:}
		(a) shows the confusion matrix of classification of lattices from the training dataset into the three classes, i.e.~stretching ($n_x<1.05$), bending ($n_x>1.95$) and the ones exhibiting combined deformation modes ($1.05 \leq n_x \leq 1.95$). Panel (b) shows the corresponding predicted confusion matrix of the trained model applied on the validation dataset. Panel (c) shows the number of lattices and the probabilities of different classes from the trained model. Panel (d) shows the corresponding predicted probabilities of the trained model applied on the validation dataset.
		\label{f:fig5}}
\end{figure*}

The resulting confusion matrix for the training and validation dataset are shown in Fig.~\ref{f:fig5}a-b. The key evaluation metric for a classification problem is the accuracy defined as, the number of correct predictions normalized by the total number of predictions. The accuracy on the training dataset converges to ~95.2\% (Fig.~\ref{f:fig5}a) and the accuracy achieved on the validation dataset is ~81.8\% (Fig.~\ref{f:fig5}b).

\begin{figure*}[!htb]
	\centering
	\includegraphics[width=0.9\textwidth]{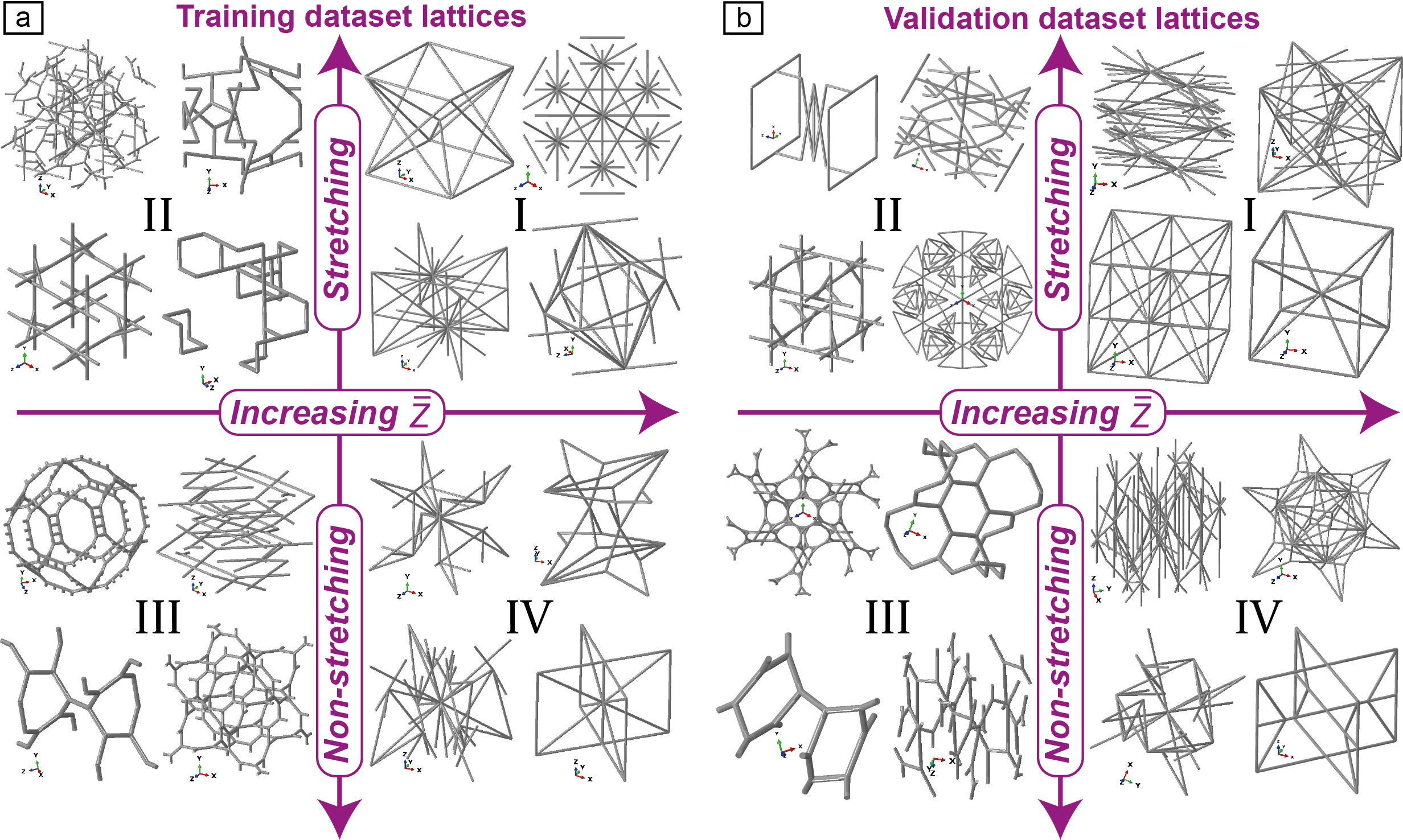}
	\caption{
		\textbf{Classified lattice visualization:}
		(a) and (b) show the lattices classified into perfectly stretching and non-stretching classes in different quadrants as a function of average connectivity, $\bar{Z}$ in the training and validation lattices, respectively.
		\label{f:fig6}}
\end{figure*}

The confusion matrix for training dataset reveals the GNN classifier only misses about ~2\% lattices in the training dataset which exhibit a perfectly stretching behavior. Out of which only ~0.5\% have a perfectly bending behavior. Similarly, the trained GNN classifier misses only ~2.2\% lattices that exhibit a perfectly bending behavior in the training dataset.

More interestingly, looking at the confusion matrix differently, the trained GNN misses only 104 lattices which exhibit a stretching behavior, i.e., only ~0.67\% of the total lattices in the training dataset. Similarly, the trained GNN misses only ~0.48\% of the lattices that exhibit a non-stretching behavior (combined and perfectly bending behavior). The resulting training accuracy for such a stretching vs non-stretching classification therefore is ~98.85\%. Accordingly, the trained GNN classifies the validation dataset comprising of 1,721 lattices into stretching and non-stretching lattices with an accuracy of ~92.27\%. This demonstrates that the trained GNN generalises very well for any lattice structure in identifying the presence of stretching mechanisms. 

The histograms of the probabilities calculated by the GNN for the different regions of the confusion matrix for the training and validation datasets are shown in Figs. 5c and d respectively. The GNN probability plots reveal that the class that differs the most from the actual class in most cases has the lowest probability and shows that the trained GNN has a high certainty.

Lastly, in Fig.~\ref{f:fig6} we inspect if the classified lattices in the training and validation dataset correlate directly with the average connectivity ($\bar{Z}$). For this purpose, the lattices in the training dataset are qualitatively placed into four quadrants, here shown with some representative examples: (i) 1st quadrant: Has a high average connectivity ($\bar{Z}$) and is perfectly stretching ($n_x<1.05$), (ii) 2nd quadrant: low $\bar{Z}$ but is perfectly stretching, (iii) 3rd quadrant: low $\bar{Z}$ and non-stretching and (iv) high $\bar{Z}$ and non-stretching. The lattices belonging to the 1st and 3rd quadrants are well recognised in the literature where connectivity dictates the dominating deformation mechanism \cite{deshpande2001}. However, the lattices belonging to the 2nd and 4th quadrants present special scenarios. They are exemplified by the complex lattices wherein a higher nodal connectivity does not necessarily deform by tensile stretching and compression of struts under uniaxial loading. This can be corroborated to the presence of a large proportion of struts which are misaligned from the uniaxial loading axis (i.e., $x$ considered here), and hence the global uniaxial loading is accommodated by bending of struts. However, a direct correlation cannot be established given the complexity of lattice architectures. Similarly, the lattices in the 4th quadrant exhibit a non-stretching behaviour although they can possess a high $\bar{Z} \geq 11$. The training accuracy of such a complex non-linear problem with hundreds of nodes and struts is obtained as ~99\% for the given dataset. The results translate to the validation dataset with nearly ~92\% accuracy and the corresponding quadrants with examples of lattices from the validation dataset are pictorially shown. The trained classifier can be then employed for any architected structure, including defects such as nodal imperfections, missing struts to predict the prevailing dominant deformation behavior along a given direction.

\vspace{5mm}
\section{Outlook}
Graph representation of lattices facilitates the construction of computational graphs in a non-Euclidean space. Such computational graphs with rich feature representation (geometrical to material) when passed into a graphical neural network optimizer yields a high accuracy on the classification problem solved here, which otherwise is highly non-linear and cannot be solved analytically. This development opens a wide range of problems, from predicting elastic moduli to non-linear behaviour (stress-strain curves) with different parent material composition. From simple lattices to multi-material and hierarchical lattices, graph based neural network can become a promising tool for researchers to designers. We anticipate trained ML models like the one presented in this work forms the basis of these future studies for many more interesting areas in architected materials beyond mechanical properties.

\vfil
\pagebreak

\bibliography{apssamp}

\end{document}